\documentclass[conference]{IEEEtran}
\usepackage{cite}
\usepackage{amsmath,amssymb,amsfonts}
\usepackage{algorithmic}
\usepackage{graphicx}
\usepackage{textcomp}
\usepackage{xcolor}
\usepackage{tcolorbox}
\usepackage[hyphens]{url}

\def\BibTeX{{\rm B\kern-.05em{\sc i\kern-.025em b}\kern-.08em
		T\kern-.1667em\lower.7ex\hbox{E}\kern-.125emX}}

\begin{document}
\title{The Emergence of Hardware Fuzzing: A Critical Review of its Significance}

\author{%
	\IEEEauthorblockN{%
		Raghul Saravanan 
	    and 
			Sai Manoj Pudukotai Dinakarrao 
	}%
	\IEEEauthorblockA{ Department of Electrical and Computer Engineering, George Mason University, Fairfax, VA, USA}%

	\{rsaravan, spudukot\}@gmu.edu

}
	\maketitle
	\thispagestyle{plain}
	\pagestyle{plain}

	
	\begin{abstract}

	In recent years, there has been a notable surge in attention towards hardware security, driven by the increasing complexity and integration of processors, SoCs, and third-party IPs aimed at delivering advanced solutions. However, this complexity also introduces vulnerabilities and bugs into hardware systems, necessitating early detection during the IC design cycle to uphold system integrity and mitigate re-engineering costs. While the Design Verification (DV) community employs dynamic and formal verification strategies, they encounter challenges such as scalability for intricate designs and significant human intervention, leading to prolonged verification durations. As an alternative approach, hardware fuzzing, inspired by software testing methodologies, has gained prominence for its efficacy in identifying bugs within complex hardware designs. Despite the introduction of various hardware fuzzing techniques, obstacles such as inefficient conversion of hardware modules into software models impede their effectiveness. This Systematization of Knowledge (SoK) initiative delves into the fundamental principles of existing hardware fuzzing, methodologies, and their applicability across diverse hardware designs. Additionally, it evaluates factors such as the utilization of golden reference models (GRMs), coverage metrics, and toolchains to gauge their potential for broader adoption, akin to traditional formal verification methods. Furthermore, this work examines the reliability of existing hardware fuzzing techniques in identifying vulnerabilities and identifies research gaps for future advancements in design verification techniques.


		
	\end{abstract}
\section{Introduction}


Addressing the intricacies of the global semiconductor supply chain demands collaborative efforts from integrated circuit (IC) designers and vendors \cite{icdesignflow1}. 
Various IC design companies or teams actively participate and contribute across the spectrum of IC design cycle, encompassing design, testing, fabrication, packaging, and integration \cite{icflow}. This collaborative engagement is integral to achieving comprehensive solutions that meet the dynamic demands of the semiconductor industry. For example, in the design and evaluation of the Apple\textsuperscript{\tiny\textregistered} 15 chip, more than 11 third-party entities have been involved in delivering sophisticated solutions for the device \cite{apple}.

Such complex modern IC design flow is vulnerable to trust issues, including the insertion of bugs and the existence of vulnerabilities due to opaqueness between different entities. 
As the IC designs are increasingly complex with heterogeneous intellectual properties (IPs) from third-party vendors, the feasibility of such bugs and vulnerabilities are 
growing at a rapid pace \cite{Artenstein'17, Liu'17, Lipp'18, Kocher'18}. For instance, in the year 2021 the number of identified common vulnerability enumeration (CVEs) recorded is close to 18,439 \cite{nistnvd, REDSCAN, mitrecve}
increased by 184\% compared to 2015 \cite{sem}. 

With the complexity of system-on-chips (SoCs) and IC designs, such increasing vulnerabilities stem from the orchestration of software and hardware components of the system. \cite{Bulck'18,Schwarz'19,Jang'16,Wojtczuk'12,Feng'22,Ghosh'23,Kang'19,Alatoun'21,Tang'22,Mohandoss'18,Vangal'08,iscas}.
For instance, Intel’s machine check bug \cite{Intel'19,Mitre} facilitates adversaries to launch a denial-of-service attack with machine check as a trigger, leading to system freeze and hanging of the processor. Re-engineering the ICs cost nearly half a Billion Dollars, along with putting Intel’s reputation at stake. 
Unlike software, hardware patches are expensive and permanent. It is crucial to rectify the CVEs and common weakness enumerations (CWEs) vulnerabilities before fabrication (i.e., pre-silicon) and release them into the market to prevent reengineering and manufacturing costs.


To nullify the effects of the vulnerabilities and bugs in the SoCs and ICs, the design verification (DV) community both from academics and industry, has proposed numerous pre-silicon DV techniques \cite{Chen'11, Mukherjee'15, Dessoky'19, Sarangi'06, Wagner'07,  Jin'05, Gogri'22, Guzey'10,kas,kas2} and Electronic Design Automation (EDA) tools \cite{Cadence, Jaspergold'23, Siemens-Questa, Cadence-Xcelium, Aldec, SymbiYosys, ABC}. It is estimated that up to 70\% of the time and effort of the IC development cycle
is spent on the verification activities \cite{Mintz'07, Spear'12, Fine'03,Wang'09, Olofsson'17}, which highlights the prominence of verification. The two popular verification methods are the dynamic \cite{Hicks'15, Sarangi'06,Wagner'07, Gogri'22 } and formal verification \cite{Cadence,Dessoky'19, Averant,Onespin,Synopsys,Siemens,Wile'05,Clarke'12, z1, zono} methods. Dynamic and formal verification methods have been used over the years for design verification and is also embedded in current-generation EDA tools \cite{Jaspergold'23, Synopsys-VCS,Synopsys,Cadence}.

However, both dynamic and formal verification \cite{ Wang'18, Tiwari'11, Ardeshiricham'17, Li'11, Li'14, Zhang'15, Meng'22   } techniques have failed to match the pace of ever-increasingly complex IC and SoC and are less efficient in detecting bugs \cite{ Trippel'22,Muduli'20,Kabylkas'21,Kande'22,Canakci'23, Qin'14}. These techniques require increased human effort, lack scalability, have limited design coverage, and the verification time exponentially increases with the design complexity \cite{Trippel'22,Kande'22,Hossain'23     }. 
Hence, there is a need for efficient and automated verification methodologies that can detect bugs compatible with the current IC design and verification flow.

To address the shortcomings of formal and dynamic verification, hardware fuzzing has been introduced in recent years \cite{Laeufer'18,Li'21, Hur'21,Trippel'22,Muduli'20, Kabylkas'21
}. Fuzzing is a widely used software testing methodology for bug detection in software applications. 
Fuzzing, in simple terms, is bombarding the software with test cases and analyzing for any invalid targets, such as memory crashes. 
Industry-based fuzzing platforms such as Google's OSS-Fuzz \cite{Serebryany'17} and Microsoft's Security Risk Detection \cite{microsoftrisk} have proved their efficacy in identifying a plethora of security vulnerabilities. In an attempt to detect vulnerabilities, Google's OSS-Fuzz has identified more than 30,000 bugs 
in 500 open-source projects \cite{Serebryany'17 }. The popular software fuzzer, American Fuzzy Lop (AFL) \cite{AFL'23}, is predominantly used in software fuzzing.  Inspired by software fuzzing \cite{Sutton'07,Bounimova'13,Nagy'19,KitPlot'15}, hardware community researchers adopted fuzzing for hardware designs \cite{Kande'22,Canakci'23,Hossain'23,Sutton'07,Bounimova'13  }. 

RFuzz \cite{Laeufer'18} is one of the first hardware fuzzing frameworks proposed in the literature. The hardware-fuzzing frameworks aim at fuzzing CPU designs for vulnerability detection. Later, some of the works proposed translating the hardware as software for adaption of software fuzzers \cite{Trippel'22, Muduli'20, Kabylkas'21}, whereas some works proposed fuzzing hardware as hardware \cite{Kande'22, Canakci'23, Hur'21} which will be discussed in Section 3.


The fuzzing entities include state-of-the-art famous RISC-V processors such as Ariane (CVA6) \cite{RISC-V}, RocketBoom \cite{RISC}, IBEX core \cite{ibex }, and peripheral components such as Advanced Encryption Standard (AES), KMAC, and RISC-V time core inspired by Google's OpenTitan framework \cite{opentitan }. 
The fuzzing frameworks have detected various hardware vulnerabilities across these platforms. In addition, some of the works have achieved increased code coverage, paving the way for deeper testing of the IC and SoC designs. 

Though the hardware fuzzing has garnered interest from researchers, multiple challenges and ambiguities still lie forward that hamper its adaptability in EDA tool flows.  
For example, the key parameters that are employed for fuzzing are the coverage metric, fuzzing engine, and evaluation metrics. However, ambiguities exist with the defined parameters and their validation. To clarify the uncertainties and pave a clear path forward for enriched hardware fuzzing technique development, 
systemization of knowledge (SoK) is pivotal. In this SoK work, 
we lay out and analyze the existing techniques in terms of their pros, cons, and ability to meet the golden standards required for their wider adaptation and standardization. 
The cardinal contributions of this SoK are: 

\begin{itemize}
	\item We lay out the fundamental principle of the Design Verification (DV) techniques and its limitations.
	\item The overlooked challenges and fundamental issues with deploying existing tool flows for hardware fuzzing is discussed. 
	\item The leniency with employing traditional verification metrics and inability to exploit full potential of fuzzers when employed for hardware fuzzing is discussed. 
	\item The delusion of hardware fuzzing in the hardware domain is outlined. 
\end{itemize}

\begin{figure*}[h]
	\centering
	\includegraphics[width=0.85\linewidth]{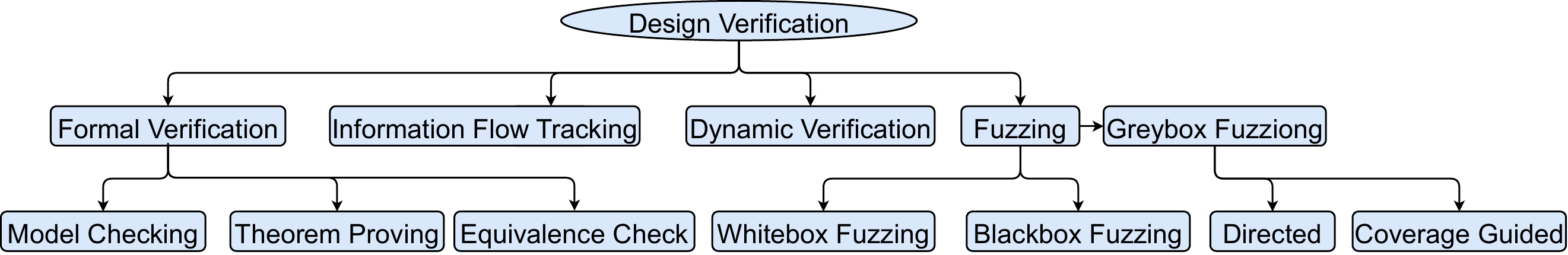}
	
	\caption{ Design Verification Classification} \label{fig:vertypes}
	\vspace{-1em}
\end{figure*}
\vspace{-0.6em}
\section{Background on Design Verification }


Hardware verification techniques for vulnerability detection can be broadly classified into {\em  1) Formal Verification, 2) Information Flow Tracking, 3) Dynamic Verification, and 4) Recently introduced Hardware Fuzzing inspired by software testing} as shown in Figure \ref{fig:vertypes}

\vspace{-0.4em}

\subsection{Formal Verification}
\vspace{-0.25em}


Formal verification is one of the conventional methods in hardware verification techniques that ensure the correctness and reliability of complex integrated circuits and electronic systems \cite{Cadence, Wile'05,Clarke'12, Hicks'15, Guo17,Guo2017,Guo3017,Farahmandi,   bmc, Drzevitzky2010,Z3,coq }.  It encompasses a diverse range of mathematical theorems and logical techniques to prove the correctness of hardware designs against specified properties. Formal verification is supported by commercial EDA tools such as Cadence Jasper Gold and Synopsys VCS Formal \cite{Jaspergold'23, Synopsys, Synopsys-VCS}. Formal verification has gained significant traction due to its ability to mitigate deep flaws in hardware design.
However, the existing formal verification techniques are crippling due to their inability to cope with increasingly large complex modern processor designs and the need for expert knowledge   \cite{Kande'22, Dessoky'19}. 
\vspace{-0.5em}
\subsection{Information Flow Tracking (IFT)}
\vspace{-0.25em}
Information-flow tracking (IFT) is a method that tracks and controls the flow of information within the hardware system for verification. It is often applied to identify and prevent unauthorized or unintended information flows, such as sensitive data leaks \cite{Tiwari'11, Ardeshiricham'17, Li'11}. 
Several secure RTL programming languages \cite{Li'11, Li'14,Zhang'15,Guo17}, 
have been created to assess non-interference properties, primarily based on IFT \cite{Guo17,Guo3017}. In this technique, the input signals are labeled, and the labels are propagated throughout the design to detect and identify design vulnerabilities, including information leakage, data manipulation, or design caveats \cite{Guo2017}. 
With the growing complexity and the size of current designs (with several thousands of lines of code), the labels often get polluted or lost, making the IFT impractical \cite{Laeufer'18,Trippel'22, Canakci'23}. 

\vspace{-1em}
\subsection{Dynamic Verification}
\vspace{-0.25em}

\begin{figure}[htb!]
	\vspace{-1em}
	\centering
	\includegraphics[width=1\linewidth]{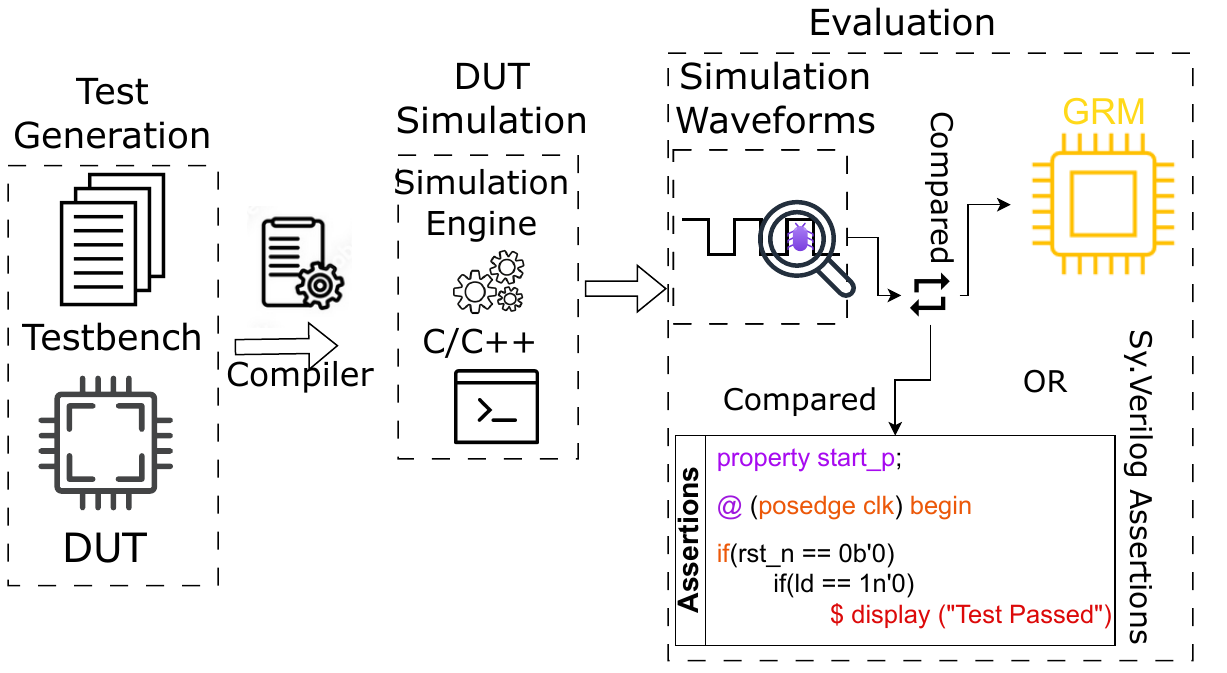} \vspace{-1em}
	\caption{Dynamic Verification
	} \label{fig:dynamic}
	\vspace{-0.6em}
\end{figure}

Dynamic Verification (DV), a.k.a runtime bug detection, is a popular verification technique. 
Dynamic hardware verification techniques simulate or emulate the design-under-test (DUT) with input seeds to obtain the outputs and analyze for bugs and vulnerabilities. 
The three cardinal steps in DV are {\em 1) Test Generation 2) DUT Simulation 3) Evaluation} as shown in Figure \ref{fig:dynamic}. 

\paragraph{Test Generation}
DV engineers craft the input test vectors during the test generation phase to simulate the DUT.  The most popular test generation techniques are the Constrained Random Verification (CRV) \cite{uvm, Ioannides, Yuan} and Coverage Directed Test Generation (CDG) \cite{Trippel'22 }. 

CRV techniques, though require human intervention, are widely practiced in DV techniques such as Universal Verification Methodology (UVM) \cite{uvm, Cocotb}. To evade the shortcomings of CRV, CDG is proposed \cite{Teplitsky2015CoverageDD, guzey }, where the input sequences are mutated based on feedback code coverage metrics to explore the non-executed regions of the design. But CDG suffers from deployment bottlenecks (i.e., inefficiency of coverage tracing) for a complex DUT \cite{Trippel'22}.

\paragraph{DUT Simulation}
Once the test vectors are generated, the HDL of the design can be simulated through various commercially available and open-source EDA tools \cite{Synopsys-VCS, Verilator, Cadence-Xcelium} for extracting the output responses. 
The simulation tools translate the HDL design to a C or a C++ program, i.e., an equivalent software model. The compiler compiles the translated software model with the testbench to provide a binary executable file (.exe) called Hardware Simulation Binary (HSB). Post compilation, the simulation engine executes the HSB with the inputs from the testbench to obtain the outputs. 

\paragraph{Evaluation}
Test evaluation is performed by monitoring the outputs of the hardware for a given input stimuli. The common methods for evaluation are to make use of the Golden Reference Model (GRM)  or System Verilog Assertions (SVA). The model is examined for any violation against the defined assertion properties or the behavior of the GRMs.  



\vspace{-0.25em}
\subsection{Hardware Fuzzing}

\begin{figure}[htb!]
	\vspace{-1em}
	\centering
	\includegraphics[width=0.9\linewidth]{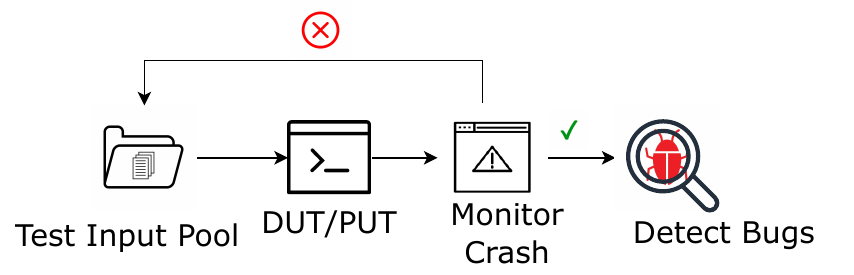} 
	\caption{Overview of Software Fuzzing 
	} \label{fig:fuzzing}
	\vspace{-0.5em}
\end{figure}

\vspace{-0.25em}
To surpass the limitations in the existing DV frameworks, hardware fuzzing \cite{Laeufer'18,Trippel'22,Kande'22,Muduli'20} has gained traction due to its popularity in the software testing community \cite{Wang'17,Bohme'17, Hongfuzz'22}. 
The concept of fuzzing primarily involves: {\em a) test generation; b) monitoring the DUT/program-under-test (PUT); and c) analyzing for bugs or errors} as shown in Figure \ref{fig:fuzzing}.

The core concept behind traditional fuzzing begins with the generation of acceptable test case generations (i.e., input stimuli). The input stimuli are then fed to the DUT/PUT for monitoring the running status of the program and is subjected to record any crash during the period of fuzzing. The monitored output is analyzed for bugs. The DUT/PUT is bombarded with test input until a crash is recorded. Software fuzzers analyze crashes for bug detection, whereas in hardware fuzzing, the expected outcome from DUT is verified against assertions or expected output from the GRM. Fuzzing incurs low deployment costs and a reduced verification period for testing complex designs.

Depending on the available information regarding the DUT, fuzzing techniques can be broadly classified into three types: i) Blackbox fuzzing, ii) Greybox fuzzing, and iii) Whitebox fuzzing. 
In black box fuzzing \cite{Libfuzzer'21}, the fuzzer does not have access to the internal details or the source code of the DUT/PUT being tested. However, the bug detection capabilities are constrained due to limited code coverage, inefficient input generation, lack of knowledge of the system's behavior, and limited feedback for debugging \cite{Libfuzzer'21}. In contrast, white box fuzzing depends upon the target code access. In Greybox Fuzzing (GF) such as AFL \cite{AFL'23}, the fuzzer has limited knowledge about the DUT while extracting the peripheral information about the data flow, control flow, data format, protocols, and high-level architecture. These fuzzers discard the primary source code after the necessary instrumentation for the program is added. The instrumented binary is subjected to monitoring based on the coverage reports.



\begin{figure}[htb!]
	\vspace{-1em}
	\centering
	\includegraphics[width=1\linewidth]{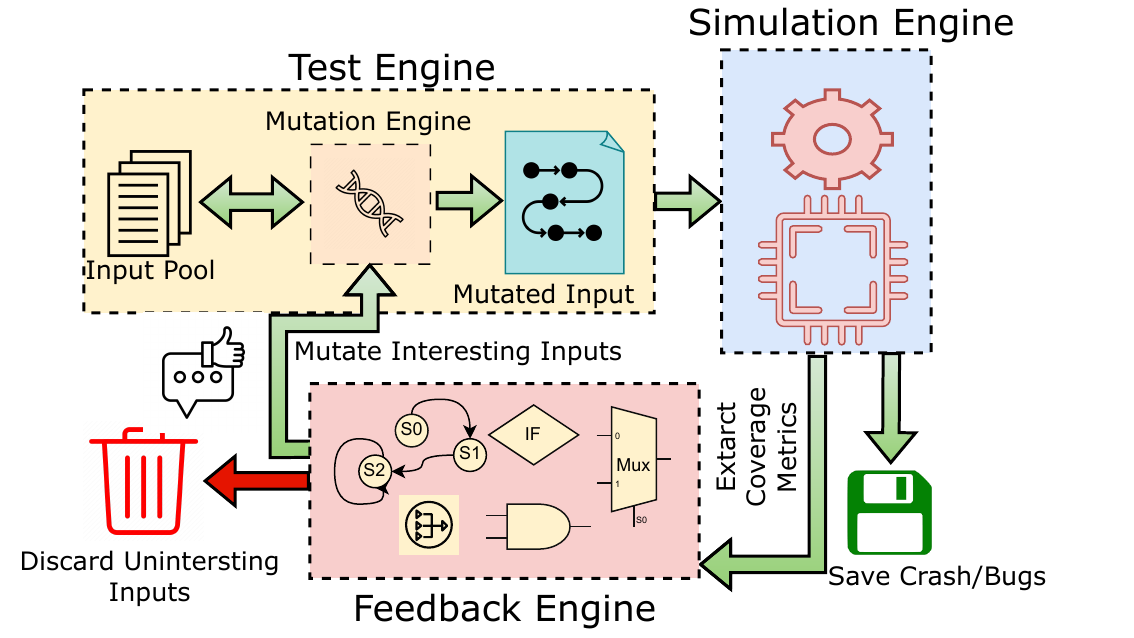} \vspace{-1em}
	\caption{Coverage Greybox Fuzzing
	} \label{fig:cfg}
	\vspace{-0.5em}
\end{figure}

Based on the code exploration strategies, fuzzers can be further classified as Coverage-based Greybox Fuzzing (CGF) and Directed Greybox Fuzzing (DGF). 


\noindent\textbf{Coverage-based Greybox Fuzzing:} In CGF, the fuzzing is aimed at achieving maximum code coverage through feedback engines (i.e., coverage metrics). In hardware, the coverage metrics that can be used are Finite State Machine (FSM), line, conditional, and MUX toggle. As shown in Figure \ref{fig:cfg}, a set of input seeds is stored and passed to the mutation engine that performs mutation operations to generate multiple input seeds. During the runtime, the coverage reports are extracted based on the input provided to the DUT and given as feedback to the mutation engine. Based on the coverage feedback, the mutation engine further mutates the interesting input seed to generate a new set of seeds. The uninteresting input seeds are discarded from the input pool. The HSB is executed with these seed inputs, and any potential crashes are saved for analyzing the vulnerabilities. 

\noindent\textbf{Directed Greybox Fuzzing:} Hardware and Software designs often undergo revisions for updating a component for better performance (i.e., incremental designs). 
To fuzz such incremental designs, DGF is used for fuzzing at a particular region rather than the whole DUT/PUT, resulting in reduced verification time.



\vspace{- 4 pt}

\section{State-of-the-Art on Hardware Fuzzing} \label{in}
This Section briefly discusses recent hardware fuzzing works as shown in Table \ref{table:fuzzing}. 

\vspace{-0.6em}
\subsection{Direct Adoption of Software Fuzzer for Hardware  }
\vspace{-0.5em}

\begin{figure*}[h]
	\centering
	\includegraphics[width=0.9\linewidth]{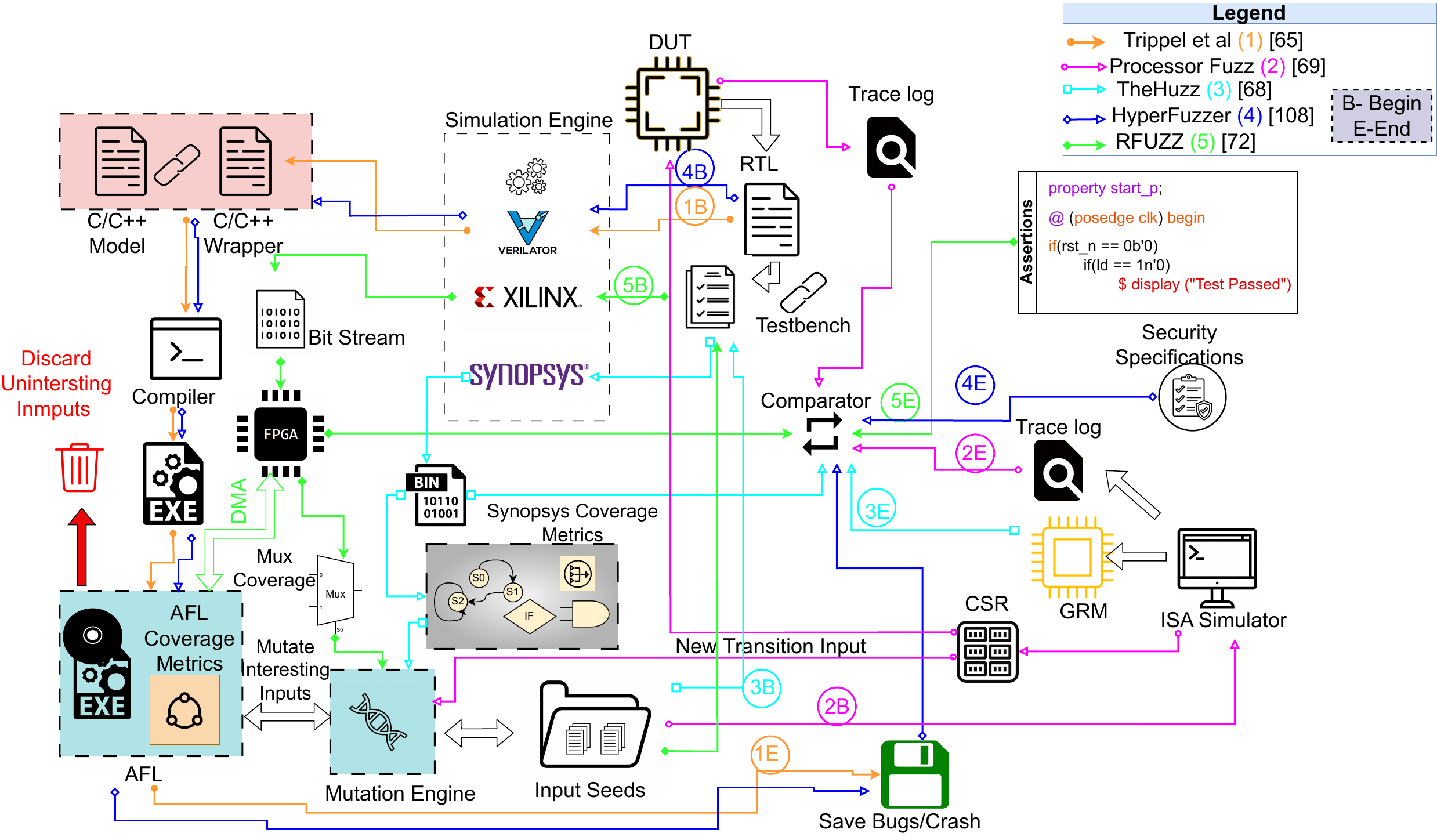}
	\vspace{-1em}
	\caption{ Hardware Fuzzing Frameworks} \label{fig:fuzzing farmeworks}
	\vspace{-1em}
\end{figure*}

As a solution to the CDG shortcomings such as deployment bottleneck  (i.e., inefficiency of
coverage tracing, 
RFuzz \cite{Laeufer'18} is the first FPGA emulation-based hardware fuzzing technique introduced as shown in Figure \ref{fig:fuzzing farmeworks}. 

The RFuzz translates the target HDL to FIRRTL \cite{Chisel} such that instrumentation is leveraged through compiler passes, enabling the test harness generator to design a wrapper for the RTL design. 
The input pins are concatenated to form a bit vector and are mapped to a series of bytes representing the input values. To ensure the tests are deterministic and repeatable, the RTL should be reset to a known test before each test execution.  {\em MetaReset} (for resetting the registers to zero for enabling  DUT reset) and  {\em Spare memories} (to reset the memory locations that have been written in previous test execution) are deployed to enable quick RTL reset without modifying the DUT characteristics. Furthermore, the instrumentation is appended to enable  {\em MUX control coverage} in hardware, ensuring FPGA-synthesizable coverage extraction. RFuzz utilizes the AFL-based mutation functions. 
The AFL fuzzer efficiently communicates with the FPGA through high-speed Direct Memory access (DMA), as shown in Figure  \ref{fig:fuzzing farmeworks}. 

\vspace{-0.5em}

\subsection{Fuzzing Hardware as a Software}
\vspace{-0.25em}

In contrast to RFuzz, Trippel \textit{et al.} \cite{Trippel'22} proposed fuzzing hardware-like software rather than porting software fuzzers directly on the hardware designs. The cardinal aspect of fuzz hardware like software is driven by the executable software version of the hardware provided by the hardware simulation tools. The translated software version of the hardware is compiled, instrumented and fuzzed using AFL for bug detection.

In \cite{Trippel'22}, the authors make use of Verilator to translate the hardware to an equivalent software model as shown in Figure \ref{fig:fuzzing farmeworks}. 
Verilator \cite{Verilator} is a popular open-source tool that leverages the equivalent software model in C++ for the given System Verilog of the RTL design. 
The C++ program can be instantiated with a C++ wrapper acting as a testbench to output HSB and simulate the device. To trace the hardware coverage in the software domain, it exploits the seamless binary instrumentation provided by the AFL fuzzer to trail code coverage. 

In addition, the verilator generates C++ code for both blocking and non-blocking statements of System Verilog and interpolates assertions and coverage points, ensuring functional coverage. With edge coverage from AFL, both code and functional coverage of the hardware in a software domain can be monitored. 
The fuzzing cores are monitored for crashes and are evaluated against SVAs.    
This work monitors for crashes such as improper memory mapping, errors in registers, buffer overflow, and floating-point computation for vulnerability detection, 
as the hardware is modeled as software.

\vspace{-0.7em}
\subsection{Fuzzing Hardware as Hardware}
\vspace{-0.5em}
Recent advancements in hardware fuzzing have embraced a domain-specific approach to eliminate the need for translating hardware to software, along with the associated preliminary tasks and equivalence checks. 
TheHuzz \cite{Kande'22}, DiFuzz \cite{Hur'21}, and ProcessorFuzz \cite{Canakci'23} adeptly integrate hardware fuzzing into conventional industry-standard hardware design and verification workflows.


\paragraph{TheHuzz}
The three pivotal components of TheHuzz include the Seed Generator, Stimulus Generator, and Bug Detector, as illustrated in Figure \ref{fig:fuzzing farmeworks}. Operating at the instruction set architecture (ISA) abstraction level, the Seed Generator furnishes the input seed in the form of instruction sequences. The Stimulus Generator then mutates the instruction at the binary level, ensuring that all bits of the instruction undergo mutation to test the processor with illegal instructions. This includes mutating opcode bits and data bits to unveil unexplored datapaths. 

The RTL design of the processor \cite{Morlkx, ibex,RISC-V}, is simulated with the binary format of the instruction using Synopsys VCS, a tool entrenched in the semiconductor industry for several decades. Synopsys VCS \cite{Synopsys-VCS} trace code coverage through various metrics, including branch, condition, toggle, FSM, and functional coverage. Based on these coverage metrics (the feedback engine), optimal weights are assigned to each instruction-mutation pair, and uninteresting pairs are discarded. Bugs are identified by simulating the ISA emulator \cite{SPIKE} (GRM), and comparing the behavior of the RTL design against the GRM behavior. 
Seamless compatibility with traditional EDA tools and requiring minimal additional effort are some of the unique aspects of this work.

\paragraph{ProcessorFuzz and DiFuzzRTL}
Other prominent works associated with hardware-specific fuzzing to detect CPU bugs are the ProcessorFuzz as shown in Figure \ref{fig:fuzzing farmeworks} \cite{Canakci'23}, and DiFuzzRTL \cite{Hur'21}. At an elevated stratum of hardware abstraction, a Central Processing Unit (CPU) manifests as an intricately designed FSM. It is indispensable to monitor these transitions of CPU states for potential bug detection for a given assembly program. ISA simulators are used to simulate the functional behaviors of the CPU, serving as a reference model for the existing fuzzing frameworks. 
Rather than using coverage metrics from the RTL simulation, ProcessorFuzz exploits CSRs from the ISA simulator as coverage metrics.
The CSR values and the transitions in them represent the changes in the architectural state flow of the CPU. 

The input pool for the ProcessorFuzz encompasses a pool of instruction at the assembly level adhering to the target ISA upon which mutation operations are performed. Rather than simulating the RTL with the mutated input, ISA simulator simulates the targeted CPU with the mutated input sequences. During the simulation, the simulator initiates trace logs from which CSR transitions, program counter, and disassembled instructions can be obtained. The trace logs of the RTL and ISA emulators are compared to detect CPU vulnerabilities. In contrast, DiFuzzRTL captures the FSM state transition of the RTL design through static analysis of the small group of registers, which plays a vital role in controlling the states of the CPU. The feedback is given through the register coverage in RTL and is compared against the ISA for evaluation of vulnerability detection.


\begin{scriptsize}
	\begin{table*}[h!]
		
		\centering
		\caption{Existing Fuzzing Framework}
		\label{table:fuzzing}
		\begin{tabular}{|l|l|l|l|l|l|l|}
			\hline
			\textbf{Framework } & \textbf{Fuzzer} & \textbf{Input} & \textbf{Simulator} & \textbf{Coverage Metric} & \textbf{Target Design } & \textbf{Comparison}    \\
			\hline
			\hline
			RFUZZ \cite{Laeufer'18}  & HW Fuzzer & Series of bits & Any & Mux Toggle & Peripherals, & Assertion\\
			&&&& &RISC-V& \\
			
			\hline
			Li et al \cite{Li'21} & HW Fuzzer  & Series of bits & PyRTL & Mux Toggle & RISC-V & Assertion \\
			&&&& &Opencore 1200& \\
			\hline
			DifuzzRTL \cite{Hur'21}  & HW Fuzzer & Assembly & Any & Register Coverage & RISC-V CPU & GRM\\
			\hline
			Trippel et al \cite{Trippel'22} & SW AFL Fuzzer & Byte Sequence & Verilator & Edge Coverage & AES,HMAC  & SW crashes \\
			&&&& &KMAC, Timer& \\
			\hline
			HyperFuzzer \cite{hyperprop}  & SW AFL Fuzzer & Series of bits & Verilator & High-level & SoC  & Assertion \\
			\hline
			DirectFuzz \cite{direct} & SW AFL Fuzzer & Series of bits & Verilator & MUX & Peripherals, & Assertion \\
			&&&& &RISC-V& \\
			\hline
			TheHuzz \cite{Kande'22}  & HW Fuzzer & Assembly  & Synopsys VCS & FSM, Branch,toggle, & RISC-V  & GRM \\
			& & & & conditional& & \\ 
			\hline
			Processor Fuzz \cite{Canakci'23} & HW Fuzzer & Assembly & Verilator & Control path& RISC-V  &GRM \\
			&&& &  register, ISA-tranistion& & \\
			\hline
			SoC Fuzz \cite{Hossain'23}  & HW Fuzzer & Byte Sequence & Xilinx ISA & Randomness, target & SoC & Database \\
			&&&& output, input coverage& & \\
			\hline
		\end{tabular}
	\end{table*}
\end{scriptsize}

\subsection{SoC Fuzzers}
\vspace{-0.25em}
In contrast to the fuzzing techniques discussed above, works such as \cite{hyperprop} focus on fuzzing SoCs. 
It employs Hyperproperties \cite{hyperprop}, which are higher-level properties that describe security policies by comparing the behavior of instances of a system. The concept of hyperproperties is commonly employed in analyzing security protocols and detecting bugs in a system. These hyperproperties can be used for laying out the SoC security specifications. Hyperfuzzing, as shown in Figure \ref{fig:fuzzing farmeworks} \cite{Muduli'20}, encompasses defining SoC security properties that adhere to confidentiality, integrity, and noninterference expressed in HyperPLTL (Hyper Past-time Linear Temporal Logic). Hyperfuzzer complies with CGF with High-Level coverage metrics. The famous Verilator is used for SoC RTL simulation, which is further instrumented to collect coverage metrics. AFL fuzzes the instrumented code to find potential bugs and is evaluated using a property checker by comparing it against the security property specifications (i.e., hypeproperties) in HyperPLTL. 

\vspace{-0.5em}
\section{Is Hardware Fuzzing a Delusion?}

\vspace{-0.25em}

The state-of-the-art fuzzing-based DV frameworks exhibit proficiency in fuzzing various IC/IP entities, 
as shown in Table \ref{table:fuzzing}. Nonetheless, these cutting-edge frameworks are not without certain limitations. 
In this section, we shed light on the often underestimated drawbacks within these domains, posited as potential impediments to the optimal efficiency of fuzzing frameworks.

\vspace{-0.5em}
\subsection{Tool Dependency}

\vspace{-0.25em}

Existing works on hardware fuzzing have chosen a wide range of tools, such as Verilator \cite{Verilator} and AFL \cite{AFL'23}, for fuzzing as outlined in Table \ref{table:fuzzing}. The limitations associated with these tools are deconstructed below.

\subsubsection{ Equivalency of Hardware Bugs}

The distinction between hardware and software lies not only in their design but also in their behavioral exhibits. 
Employing software fuzzers directly onto the RTL poses its own set of challenges. In the software domain, a common definition for a bug is an outcome marked by crashes, signifying an improper functioning of the operating system, which in turn crashes the fuzz engine. However, such scenarios do not exist in the hardware realm. 

\vspace{-0.5em}
\begin{tcolorbox}
	Limitation A.1.1: AFL fails to report hardware bugs, as crash or hang doesn't exist in hardware. Hardware inherently does not crash; instead, it produces erroneous outputs.
\end{tcolorbox}
\vspace{-0.5em}

For instance, a bug in a software application that causes it to access invalid memory locations, such as CVE-2023-3953 \cite{CVE'3953}, can be detected as a crash by AFL. 
On the other hand, hardware, instead of crashing, issues often manifest as malfunctions or failures in the form of faulty outputs. Because of this limitation,  techniques such as RFuzz  \cite{Laeufer'18}, which relies on employing AFL directly on the hardware, fail to detect any hardware vulnerabilities. 

\subsubsection{Dependency on HDL and its Translation} 

Fuzzing techniques such as Rfuzz \cite{Laeufer'18} translate the given RTL to FIRRTL \cite{Chisel}, casting dubitable shadows over their equivalence. In addition, RFuzz is based on Midas, lacking language compatibility with FIRRTL. This support is currently confined to Chisel, Verilog, and specific segments of System Verilog \cite{firrtl}. These factors oblige the users to have the expertise and prior knowledge of the languages, fostering fuzzing difficulties. 

\vspace{-0.5em}
\begin{tcolorbox}
Limitation A.2.1: FIRRTL is limited to only certain languages that require expertise and prior knowledge.   
\end{tcolorbox}
\vspace{-0.5em}


To exhibit the shortcomings of deploying software fuzzers directly on RTL designs, hardware designs are translated to software models such as binary executables on which AFL is deployed. Such translation introduces additional challenges. 
Verilator \cite{Verilator} is a popular open-source tool that leverages the equivalent software model in C++ for the given System Verilog of the RTL design. The RTL design is subjected to parsing and lexical analysis to format the Abstract Syntax Tree (AST) \cite{GNU'94, ast} from which the
verilator generates an equivalent C++ of the hardware.

However, there arises a question on the equivalency be-
tween the hardware and the translated hardware. The ver-
ilator fails to capture inherent hardware behaviors such as
signal transitions, FSMs, and floating wires described in
HDL languages to software constructs. In addition, the
verilator requires linking auxiliary libraries to simulate
the translated hardware, generating an emulated model
rather than a simulated model. A translated hardware design needs to consider the fundamental hardware behaviors, including bus transactions, register computation, and controllers. These aspects constitute the foundation for system operations. 
Limitation A.1.1, along with A.2.2, is well proven by the results of employing AFL directly on RTL \cite{Laeufer'18}  and translating hardware as a software \cite{Trippel'22}, which failed to report hardware bugs, whereas the later works \cite{Hur'21} reported the bugs. 

\begin{tcolorbox}
Limitation A.2.2: Tools such as Verilator fail to capture inherent hardware behaviors
such as signal transitions, FSMs, and floating wires
described in HDL languages to software constructs.
\end{tcolorbox}
\vspace{-0.5em}



\subsubsection{ FPGA Overheads}

FPGA-based fuzzing has been introduced in \cite{Laeufer'18} to accelerate the fuzzing and 
minimize the software conversions. However, FPGA-based fuzzing also 
requires traditional software fuzzers such as AFL, which 
introduces limitations such as A.1.1. 
In addition, FPGA for accelerating fuzzing incur fuzzing costs, and complex designs may not map to FPGA fabric structures, promoting scalability constraints. Fuzz testing on FPGAs can be computationally intensive and time-consuming \cite{CWE-400}. Hardware encapsulates the physical components of a computing system, encompassing tangible entities such as processors, memory modules, and peripheral devices. Simulating these complex, intricate hardware blocks requires a substantial amount of time than the translated hardware model (i.e., the software model of the hardware). For instance, fuzzing of Ariane CPU, which is capable of hosting LINUX OS, has 20,968 lines of code (LoC) \cite{ariane}. Generating a large number of random test vectors for complex CPU designs leads to memory and resource overheads. Fuzzing FPGAs may not be easily portable to different designs or platforms, as they are often customized for the target board(s) \cite{wong}.  


\vspace{-0.5em}
\begin{tcolorbox}
Limitation A.3.1: Fuzz testing on FPGAs can be computationally intensive and time-consuming. 
\end{tcolorbox}
\vspace{-0.5em}

\subsubsection{Instrumentation Overheads}
Obtaining coverage metrics is crucial in fuzzing, for which instrumentation is performed. 
As the majority of the fuzzers, including AFL, are inherently software fuzzers, the instrumentation supports only software models, 
which limits to deployment of AFL directly for hardware designs. 

\vspace{-0.5em}
\begin{tcolorbox}
Limitation  A.4.1: Instrumentation provided by AFL does not support HDL constructs. 
\end{tcolorbox}
\vspace{-0.5em}


Even though efforts such as translating the hardware to software were taken to circumvent these challenges \cite{Trippel'22},  Limitation A.2.2 still persists and lacks code coverage, scalability for complex CPU designs, and instrumentation overheads. For example, instrumenting an Ariane Processor (20,968 LoC) leads to large instrumentation overheads corresponding to (CWE-400) \cite{Kande'22, CWE-400}.
Besides these, the primary difference between hardware and software is in the definition of input arguments. The input for the software fuzzing is in the form of bytes, whereas for the hardware, it is in the notion of input pins accepting values at each clock cycle. The input format for a translated hardware design should be well crafted in order to perform meaningful mutations and effective fuzzing.

\vspace{-0.5em}
\begin{tcolorbox}
Limitation  A.4.2: Unoptimized instrumentation leads to significant instrumentation overheads - CWE-400. 
\end{tcolorbox}
\vspace{-0.5em}

\vspace{-0.5em}
\subsection{Reliance on Coverage Metrics}
\vspace{-0.5em}

The existing works on hardware fuzzing and hardware verification have chosen a wide range of coverage metrics to detect functional and security vulnerabilities, as outlined in Table \ref{table:fuzzing}.

\subsubsection{Limitation of Verification Capability due to Chosen Coverage Metric(s)}

The capability of hardware to detect a vulnerability highly depends on the considered coverage metrics. 
For instance, RFuzz employs MUX toggling as the coverage metric to determine whether an input seed is interesting or not. 
This captivates the RFuzz to capture the RTL design implementations of MUX expressed in combinational logic impending to account for the coverage point \cite{Kande'22}. Thus, RFUZZ will not cover any combinational logic that does not drive the select signals of the MUXes. 



\vspace{-0.5em}
\begin{tcolorbox}
Limitation B.1.1: Selection of coverage metrics is prone to human bias, limiting verification capability.
\end{tcolorbox}
\vspace{-0.5em}

ProcessorFuzz \cite{Canakci'23} utilizes CSR as a coverage metric to detect bugs. 
Use of CSR-based metrics leads to the detection of Read-after-Write dependencies vulnerabilities, which may not be 
detected using MUXes. Furthermore, 
if semantic correctness and timing information are considered, then the possibility of detecting such attacks is viable. 
Hence, meticulous planning is required for selecting the coverage metrics as it is influenced by humans, which is often prone to biased selection.



In addition to the manual selection of metrics, fuzzers such as AFL 
utilizes the in-house code and edge coverage metrics to trace the fuzzing efficacy on the translated hardware design \cite{AFL'23, Trippel'22}. However, extracting actual hardware coverage metrics from the translated design is questionable as there exists a fundamental difference between hardware and software coverage metrics. The software coverage metrics line and edge coverage is comparable in some hardware instances, however, other coverage metrics, such as FSM or MUX Toggle are not transferrable. These coverage metrics are required for detecting vulnerabilities associated with hardware \cite{Kande'22}.

First of all, these software coverage metrics are valid if and only if the limitation A.1.1 and A.2.2 is resolved. As there is a disparity between the RTL design and the translated hardware design, usage of inbuilt AFL coverage metrics is not viable and inefficient failing to detect CWE-705 \cite{CWE-705} as well.

\vspace{-0.5em}
\begin{tcolorbox}
Limitation B.1.2: AFL coverage metrics doesn't imply to hardware and fails to detect CWE-705.
\end{tcolorbox}
\vspace{-0.5em}

\subsubsection{Robustness of Coverage Metrics} 

To bypass the limitations existing with the software coverage metrics and to account for hardware coverage metrics, as mentioned earlier, some of the hardware fuzzing frameworks (i.e., fuzzing hardware as hardware) utilize the traditional hardware coverage metrics provided by the EDA tools. These frameworks extract the line, code, FSM, toggle, and branch coverage to capture the behaviors of the RTL design. However, RTL simulations for CPU designs are slower compared to ISA simulations. ISA simulations are 75$\times$ faster than RTL simulations \cite{Canakci'23}.
With these advantages, some works use the CPU's Control Status Registers (CSRs) to steer the fuzzing. However, multiple robustness challenges exist while capturing the coverage metrics in a reliable manner. 
For instance, the reliability challenges associated with CSR are outlined below. 

\vspace{-0.5em}
\begin{tcolorbox}
Limitation B.2.1: Reliable extraction of coverage metrics is challenging.
\end{tcolorbox}
\vspace{-0.5em}

\paragraph{Limited Granuality}  CSRs often provide coarse-grained coverage information. They may not offer fine-grained details about which specific portions of the hardware design have been tested or remain untested. This lack of granularity can make it challenging to pinpoint precise areas that need additional testing or improvement.

\paragraph{Difficulty in Identifying Root Causes} CSR-based coverage metrics may not assist in identifying the root causes of issues. They can indicate whether certain control paths have been exercised but might not provide insights into why failures occurred or how to address them effectively. These fail to identify side-channel attack CVEs on the processors \cite{Canakci'23}. 

\paragraph{Architecture-Specific} CSRs are architecture-specific. Each processor or system may have its own set of CSRs with unique features and behavior. This can make it challenging to write code that is portable across different platforms. For example x86 processors  have a variety of control registers, such as CR0, CR2, CR3, and CR4 \cite{csrx86}, whereas RISC-V uses mstatus, misa, mip \cite{csrrisc}

\paragraph{Complexity and Overhead} Implementing CSR-based coverage tracking can add complexity and potential overhead to the testing process. This may result in increased computational resource usage and potentially slower testing.

\vspace{-0.5em}
\begin{tcolorbox}
Limitation B.3.1: The majority of the coverage metrics primarily focus on functional checks but not parametric behavior (such as temporal behavior).
\end{tcolorbox}
\vspace{-0.5em}

\subsubsection{Can Coverage Metrics Capture Hardware Bugs or
Vulnerabilities?}

The existing works on hardware fuzzing as outlined in Table \ref{table:fuzzing} have chosen
a wide range of coverage metrics to detect functional and security vulnerabilities.
Although usage of these metrics can detect functional bugs, the security vulnerabilities such as buffer overload  CVE-2023-29856  \cite{CVE'23}, data leakage CVE-2017-5927 \cite{CVE'17},  CVE-2023-32342  \cite{CVE'23-2}  can still be left undetected through these metrics. For instance, though the width of input ports match the incoming signals, timing violation can lead to erroneous inputs and outputs, which is analogous to the CVE-2023-29856 bug in the D-Link hardware \cite{CVE'23}.

Consideration of these non-parametric metrics
will lead to the detection of exploits including \cite{CVE'04,CVE'21a, CVE'17, CVE'23-2 }.
Though DiFuzz \cite{Hur'21} facilitates the identification of side-channel vulnerabilities, it requires prior knowledge of the details of microarchitectural features. Furthermore, it requires static analysis to monitor a small group of registers in the RTL design. A CPU is a complex system encompassing thousands of memory elements. It is tedious and error-prone to manually compare the registers in the RTL designs against the registers in the ISA emulator owing to monitoring and analyzing overheads.  

\begin{tcolorbox}
Limitation B.3.2: Detection of security vulnerabilities is often challenging due to limited information captured through coverage metrics. 
\end{tcolorbox}


\subsection{Analysis and Verification of Vulnerabilities in Hardware Fuzzing}

The pivotal component of sophisticated fuzzing frameworks lies in the meticulous evaluation criteria employed, where the fuzzer scrutinizes and contrasts the behavioral aspects of the hardware. The fuzzing frameworks exploit the usage of assertions, Golden Reference Models (GRMs) (i.e., physical hardware or GRMs through ISA), and a database of vulnerabilities as outlined in Figure \ref{fig:fuzzing farmeworks} and Table \ref{table:fuzzing}. Though these metrics evaluate the presence of bugs in the CPU, many alarming concerns are disregarded during the evaluation process.

\begin{tcolorbox}
Limitation C.1.1:  Insertion of assertion requires
manual instrumentation, which may not suffice for large
complex CPU designs as it lead to instrumentation
overheads.
\end{tcolorbox}

\subsubsection{Reliance on Assertions}
DV engineers inject System Verilog assertion (SVA), treating them as equivalent versions of software crashes to monitor any potential bugs in the code. However, to define the assertion it requires manual instrumentation, which may not suffice for large complex CPU designs as it may lead to instrumentation overheads. As mentioned earlier, hardware does not inherently crash, which fosters the question of equivalency between the SVAs and a software crash. Instead commercially available tools such as JasperGold \cite{Jaspergold'23} or Synopsys VCS \cite{Synopsys}  can be used to fuzz in the same manner with minimal instrumentation overheads and less preliminary setups. 
In contrast to software, as hardware does not offer any hang, crash, or termination, the fuzzer may not capture any potential vulnerabilities.

\vspace{-0.5em}
\begin{tcolorbox}
Limitation C.1.2:  System Verilog Assertions (SVA) and software crashes are not equal.
\end{tcolorbox}
\vspace{-0.5em}

Moreover, verification engineers predetermine all assertions, directing the fuzzers to specifically examine predefined conditions rather than uncovering unknown vulnerabilities. As the verification engineers are responsible for crafting the assertion, these predefined conditions are prone to human biases. Moreover, complex CPU designs encompassing several thousands of LoC require increased human intervention and instrumental overheads. Even though the fuzzer engine alerts against the predefined assertions, it does not analyze the underlying reason for the caused bug. These are the two cardinal underlying reasons why neither adoption of software fuzzers on hardware nor fuzzing hardware-like software has reported any potential crashes or bugs \cite{Laeufer'18, Trippel'22}. Unlike software, CPUs are complex multi-cycle systems that have an impact on the sequence of states of instructions. It may take several clock cycles to execute, decode, and commit the changes at the architectural level impending the assertions to capture the cycle of events. All these factors limit the scalability of the assertions as an evaluation metric upon which the fuzzer can rely. 

\vspace{-0.8em}
\begin{tcolorbox}
Limitation C.1.3:  System Verilog Assertions (SVA) are prone to human bias and can only examine
pre-defined conditions rather than uncover unknown vulnerabilities.
\end{tcolorbox}

\vspace{-0.5em}
\subsubsection{Reliance on GRMs} Golden Reference Models (GRMs) in hardware refer to meticulously crafted and validated models that serve as benchmarks or standards for the expected behavior of a hardware design. These models are considered the golden or authoritative representations against which the actual hardware implementation is compared during verification and testing processes. Predominantly, most of the fuzzers \cite{Kande'22,Canakci'23} rely on GRMs to trace the bug in the RTL design. The possible GRMs can be dissected into a physical hardware model or a GRM generated by curated software simulators, which poses additional reliance challenges, as discussed in Section 3.

\vspace{-0.5em}
\begin{tcolorbox}
Limitation C.2.1:  The availability of third-party  GRMs and access to the source code and internal signals are limited.
\end{tcolorbox}
\vspace{-0.5em}

The incorporation of third-party IPs along with CPU is a common practice in modern IC design methodologies, delivering exceptional real-world solutions. However, with the current generation ICs becoming complex and adopting third-party IPs \cite{Feng'22,Ghosh'23,Kang'19}, the availability of GRMs and access to the source code and internal signals are limited. This impedes the efficiency and efficacy of hardware fuzzing. 

What is more, although GRMs can be used to detect functional errors, the detection of security exploits is error-prone as they can be embedded unintentionally or can even be a side effect. The unavailability of GRMs should not impede the fuzzing, as verification of the hardware is pivotal from security and reengineering cost perspectives. For instance, CPUs have an ISA simulator \cite{SPIKE,RISC}, but a standalone IP block/peripheral unit does not have any equivalent entity to identify as a GRM. The verification engineers rely upon the RTL code, which might also have vulnerabilities. This hardware might have injected trojans, infecting the entity to which it is connected.  Fuzzing based on a buggy GRM as an evaluation criterion may lead to poor fuzzing standards and mislead the verification community with false positives.

\begin{tcolorbox}
Limitation C.3.1:  The disparities in ISA simulation tools
might result in faulty evaluations against the RTL design, limiting the bug detection capabilities.
\end{tcolorbox}

\begin{figure*}[h]
	\centering
	\includegraphics[width=1\linewidth]{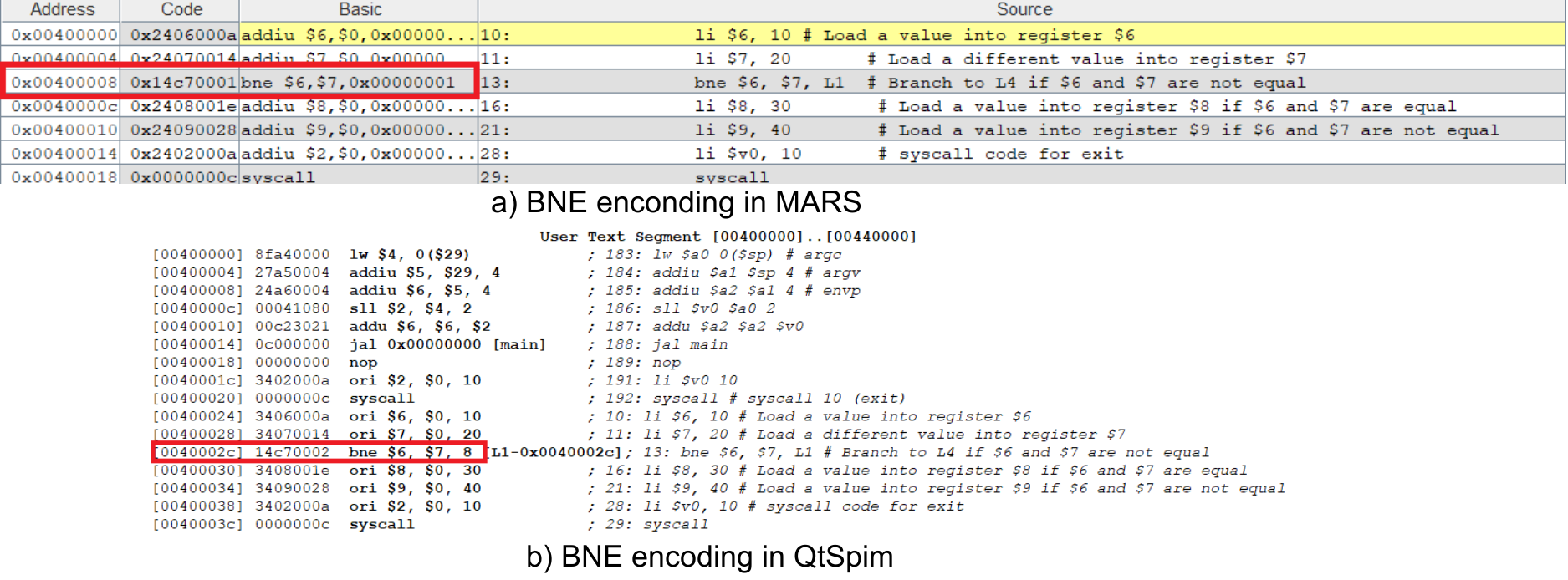}
	\vspace{-1em}
	\caption{BNE encoding in a) MARS , b) QtSpim } \label{fig:qtspimmars}
	\vspace{-1em}
\end{figure*}

 \subsubsection{Software Models}
The other way to depend on the GRMs is through software models. The CPUs can be simulated in ISA simulators (i.e., software simulations) \cite{RISC, SPIKE} to verify the functionality of the processors. For instance, the RISC-V community has crafted ISA simulators for RISC-V processors. However, these software are designed by developers and are still prone to human errors and bias.

Consider a simple case study shown in  Figure \ref{fig:qtspimmars}, which demonstrates the disparity in the encoding of the \verb|BNE| instruction in MIPS simulator \cite{mips,mars,qtspim }. We consider two popular 
ISA simulation tools, QtSpim \cite{qtspim} and MARS \cite{mars} for ISA GRM generation. 

There is inconsistency with the branch distance calculation between the outputs of QtSpim \cite{qtspim} and MARS \cite{mars} simulators, specifically related to the encoding of the \verb|BNE| (branch not equal) instruction (Line 9). The above code loads values in register location \$6 and \$7, and if the register values are not equal, it jumps to branch L4 (Line 14) using the \verb|BNE| instruction.

The \verb|BNE| instruction is of type I (immediate), and the last 16 bits are the branch distance (in words) computed from where the PC is, which is pointing to L1 after fetching the \verb|BNE| instruction. QtSpim encodes it as \textbf{0x14c70002} as shown in Figure \ref{fig:qtspimmars}, indicating "2 words ahead" or "2 instructions ahead"  from the current instruction. However, according to the standard interpretation, the branch distance should be calculated from the current instruction's location, considering it as 0 instructions ahead. Therefore, the correct branch distance from the current instruction should be 1, not 2.

In contrast, MARS encodes it correctly as \textbf{0x14c70001} as shown in Figure \ref{fig:qtspimmars}. This deviation in the branch distance calculation may suggest a potential inconsistency or discrepancy in the QtSpim simulator's handling of \verb|BNE| instruction. Some of the hardware fuzzing works such as TheHuzz \cite{Kande'22} solely depend on the instruction encoding to generate input seeds and mutate seeds. These disparities in ISA simulation tools might result in faulty evaluations against the RTL design, limiting the bug detection capabilities. Additionally, such disparities may lead to false alarms in bug detection and incur re-verification costs and delays.

Even though the GRMs are carefully curated to eliminate bugs, a research group recently found a bug associated with  incorrect memory accessing in the RISC-V cores \cite{Tricheck}.The existing hardware frameworks  {\em  TheHuzz} and {\em ProcessorFuzz} solely depends on the ISA simulator for the evaluation process. In addition, there exist bugs in the x86 ISA as well \cite{Domas2017BreakingTX}.    However, the open-source development community \cite{RISC-V} is consistently committed to rectifying the reported faults to design robust architectures for future computing needs. 

 \subsubsection{Portability Challenges due to Software Limitations}
The ISA tools are instrumented to obtain coverage metrics from CSRs. However, the instrumented ISA tool is not compatible with other processor designs following different instruction sets. For instance, the instrumented RISC-V ISA tool is incompatible with x86 ISA processors \cite{Domas2017BreakingTX}. Though the existing frameworks claim that they can be ported to other ISA targets, these claims are not proven due to the differences in their instruction sets and rules \cite{Domas2017BreakingTX,RISC-V}. 
With such discrepancies associated with ISA as an evaluation benchmark, it is not feasible to port the ISA simulator to other architectures.

\vspace{-0.5em}
\begin{tcolorbox}
	Limitation C.4.1:  The ISA tools can be used only for the processors of the same ISA target.
\end{tcolorbox}
\vspace{-0.5em}


\vspace{-0.5em}

\subsection{Fuzzing Entitites}
\vspace{-0.25em}

The current fuzzing frameworks are limited to fuzz either a CPU or a standalone peripheral IP or a SoC. None of the fuzzing frameworks provide flexibility to be adaptable across the spectrum of IC designs. 
For instance, Tripple et al. work \cite{Trippel'22} is limited to fuzz-only OpenTitanCores, whereas TheHuzz, ProcessorFuzz, and Diffuz are limited to fuzz processor designs. Furthermore, within the processor fuzzers, there exists a limit to fuzz only specific ISA associated with the proposed frameworks. TheHuzz evaluates only the functional behavior of the CPU but does not monitor temporal behaviors. This might lead to the inability to detect side-channel vulnerabilities such as specter and meltdown attacks in the processor designs.  It is indeed to evaluate the peripherals interfaced to the CPU as they may pose a huge threat as well \cite{Bartley,Zippa, HardwareSprint}. There is a need for a unanimous fuzzing framework to fuzz all the entities associated with hardware design. 

\vspace{-0.5em}
\begin{tcolorbox}
	Limitation D.1.1: Unanimous fuzzing framework for RTL/CPU Designs does not exist.
	
\end{tcolorbox}

\vspace{-1em}

\section{Conclusion}
\vspace{-0.5em}

This SoK paper provides insights into hardware fuzzing techniques for identifying bugs and vulnerabilities. We comprehensively reviewed and explained the fundamental principles of existing hardware fuzzing, the methodologies involved, and the diverse hardware designs
in which it can be employed. We have laid out the challenges/limitations existing with the current frameworks. We have given insights on how the fuzzing frameworks have been assumed and pointed out its pros and cons. The overlooked challenges and fundamental issues with deploying existing tool flows for hardware fuzzing is discussed. We also provide future possible directions for further advancement of design verification techniques.

	\bibliographystyle{IEEEtran}
	\bibliography{ref.bib}
	
	
\end{document}